# T-A Formulation to Model Electrical Machines with HTS Coated Conductor Coils

Tara Benkel, Mayraluna Lao, Yingzhen Liu, Enric Pardo, Simon Wolfstädter, Thomas Reis, Francesco Grilli

*Abstract*—Modelling high temperature superconductor (HTS) motors remains challenging mainly due to the high aspect ratio of these conductors but also because of the properties of the magnetic materials. This paper presents a 2D time dependent model to assess the AC losses of superconducting motors based on the new *T-A* formulation, which by using Finite Element Methods (FEM), allows its implementation in commercial software. The *T-A* formulation computes the magnetic flux density with different Maxwell's equations depending on the areas of the motor and makes it possible to use the thin strip approximation i.e. the HTS tapes are modelled as infinitely thin lines. The model is then expected to tackle the high aspect ratio of the HTS as well as decreasing both the mesh complexity and the computing time. The first objective of the paper is to validate the method in 2D by evaluating the AC losses of a specific synchronous motor called SUTOR; the computed results are compared with good agreements to those assessed with the MEMEP method, already validated. In a second part, the same losses are computed, taking into account the anisotropy of $J_c$ with the implementation of a data set based on experimentally measured $I_c$ at 65 K and 77 K.

*Index Terms*—HTS Machine, AC losses, *T-A* formulation, Finite Element Methods (FEM), $J_c$ anisotropy.

## I. INTRODUCTION

THE recent and continuous improvements of REBCO coated conductors (CC) [1] consistently increase their interest for electric power applications, such as energy storages (SMES) [2], cables [3], fault current limiters [4], high field magnets [5], [6], accelerators [7] and machinery [8].

More particularly, the high current density under high field of the High Temperature Superconductors (HTS) at low temperature [9], [10] is a promise for electrical machines to reach higher power densities by operating with higher currents and/or magnetic flux densities [11]. HTS machines are then expected to show both a reduction in size and weight compared to conventional ones, enhancing their attractiveness for embedded applications such as ship and aircraft propulsion for which compactness and lightweight are of paramount importance [12].

Even if HTS materials do not show losses while operated in DC, their AC losses have to be investigated as they will increase the cooling system and so decrease the overall efficiency of the machine. Nevertheless, modelling superconducting machines remains challenging and time consuming mainly due to the high aspect ratio of the HTS but also because of the properties of the magnetic materials present in electrical machines.

Recently, a hybrid model using the *A*- and *H*- formulations for modelling superconducting electrical machines has been proposed [13]. It is based on the idea of separating the model of an electrical machine in two parts, where the magnetic field is calculated with the most appropriate formulations: the *H*-formulation in the part containing the superconductors, because it handles well their non-linear electrical resistivity; and the *A*-formulation in the part containing other materials, because it handles well conventional conductors and permanent magnets. The main challenge is to determine and to correctly apply the continuity conditions on the boundary separating the two regions. Depending on the location of such boundary – in the fixed or rotating part of the machine – the conditions that one needs to apply are different.

This paper presents a new 2D time dependent model developed to tackle the challenges of modelling superconducting machines with coils made of HTS coated conductor and assess their AC losses within the same model. The model is entirely computed using Finite Element Method (FEM), which allows its whole implementation in commercial software, here COMSOL Multiphysics. It is based on the *T-A* formulation of Maxwell's equations, introduced by H. Zhang et al. in [14], which is briefly recapped in the first part of the paper. Whereas the *A-H* formulation uses different formulations for different regions, in the *T-A* formulations all regions (including those with superconductors) are simulated with the *A*-formulation; the *T*-formulation is used to calculate the current density in the superconductor domains and to pass this value back to the *A*-formulation. This approach allows an easier implementation in numerical models for electrical machines based on the *A*-formulation.

The objective of the paper is to validate the use of the *T-A* formulation to simulate superconducting electrical machines

T. Benkel, Y. Liu and F. Grilli are with the Karlsruhe Institute of Technology, Karlsruhe, Germany.
M. Lao was with the Karlsruhe Institute of Technology, Karlsruhe, Germany.
E. Pardo is with the Institute of Electrical Engineering, Slovak Academy of Sciences, Bratislava, Slovakia.
S. Wolfstädter, T. Reis are with OSWALD Elektromotoren GmbH, Miltenberg, Germany.
This work is funded by the European Commission Grant No 723119 (ASuMED).

with a 2D time-dependent model. For this purpose the model is used to calculate the AC losses of a superconducting motor, called SUTOR, presented in [15]. The results are then compared to those computed in [16] by E. Pardo et al., using the Minimum Electro-Magnetic Entropy Production (MEMEP) method [17] to model superconductors combined with conventional static FEM for the overall motor. This method has already been validated in [17] using a comparison with analytical results on straight conductors and with experimental results on pancakes and stack of them. More details can be found in [18].

In the last part of the paper, experimental measurement data of $I_c$ according to the background field components are implemented in the model for 65 K and 77 K.

## II. MODEL DESCRIPTION

The *T-A* formulation was introduced by H. Zhang et al. in [14], where the model was validated in the case of a thin disc magnetization by comparing the assessed results to the analytical ones. The model was also validated for a large number of HTS tapes against the well-established *H*-formulation in [19] and it has already been applied on stacks of REBCO coated conductors, coils [19] and HTS cables such as Conductors on Round Core (CORC) and Twisted Stacked-Tape Conductors (TSTC) [20].

### A. *T-A formulation*

As suggested by its name, the *T-A* formulation requires the implementation of both the *A* and the *T* formulations. The current density *J*, defined as an input, allows to assess the magnetic flux density, *B*, using two of Maxwell's equations depending on the areas of the motor. In the non-superconducting region, *B* is computed by evaluating the state variable *A*, the magnetic vector potential, using the Maxwell Ampere law. Finite-element method models based on the *A*-formulation are routinely used to design electrical machines, e.g. to determine the peak current in the stator windings or the generated torque. For this purpose, programs like Ansys Maxwell [21], Comsol Multiphysics [22], and FEMM [23]. The focus of the present work is the validation of the use of the *T-A* formulation on electrical machines, with the purpose of studying the electrodynamic behaviour and calculate the AC losses in the superconducting parts.

In the superconducting region, *B* is evaluated by solving the state variable *T*, which is the current vector potential related to *J* by definition.

The electric field, *E* is thereafter computed using the *E-J* power-law in which $E_c$ is the arbitrary electric field threshold usually chosen to be $10^{-4}$ Vm$^{-1}$ [24]. $J_c$, the critical current density is initially chosen to be constant, in order to validate the *T-A* formulation. Then, in the second part, its angular dependence on the magnetic field is considered, using data from experimental $I_c$ measurements.

Eventually, knowing the electric field *E*, *B* is computed using Faraday's law.

In the following, only the superconducting layer is considered in the model of the superconducting tape. The others layers can be neglected, because they do not contribute to the shaping of the magnetic field and to the total AC losses [25], [26].

### B. *Thin strip approximation*

Moreover, as long lengths of REBCO are only provided in a tape shape, the assumption can be made that the magnetic field density on the thickness of the conductor is homogeneous. The HTS tape is then modelled as an infinitely thin line. This approximation was first used to study the eddy currents on thin plates [27], [28] and has been successfully applied on HTS in 2D [29] and 3D [30], [31], [32]. The *T*-formulation makes it possible to use this hypothesis, as shown both in 2D [19] and 3D [14]. Therefore, the present model is expected to decrease both the mesh complexity and the computing time.

## III. APPLICATION OF THE METHOD

The *T-A* formulation is applied to simulate a 3-phase superconducting torque motor (SUTOR), which is a synchronous motor with rare earth permanent magnets in the rotor and whose stator windings are wound with REBCO tapes. These latter are at low temperature (65-77 K), whereas the rotor is at room temperature. The magnetic material used both in the stator and the rotor is the Cobalt Steel Vacoflux 50 [33], whose properties are already implemented in COMSOL's materials database.

The main characteristics of the motor are summarized in Table I and more information is given in [15]. As mentioned before, the aim of the current paper is to use the *T-A* formulation to simulate superconducting electrical machines and to the calculate AC losses in the superconductors. In this context, the SUTOR motor is used as a case study for testing and validating the model.

TABLE I
MAIN PARAMETERS OF THE SUTOR MOTOR

| | |
|---|---|
| Number of teeth | 12 |
| Number of pairs of poles | 4 |
| Number of slots per pole and phase | 0.5 |
| Number of coils | 12 |
| Active length | 240 mm |
| Outer diameter | 288 mm |
| Remanence of permanent magnets | 1.2 T |
| Number of turns per coil | 52 |
| Width of the superconductor | 4 mm |
| Peak current in one tape | 45.96 A |
| Nominal electrical frequency | 44 Hz |
| Nominal mechanical speed | 665 1/min |
| Nominal mechanical torque | 575 Nm |

The geometry of the SUTOR motor is illustrated in Figure 1, which also shows a typical distribution of the magnetic flux density on the motor's cross-section.

In the analysis that follows in the paper, the focus is on the two half-coils squared by a dashed red line in Figure 1, where the top half-coil and the bottom half-coil are respectively named HC6 and HC7. The field distribution illustrated in Figure 1 is computed when HC7 carries its maximal amplitude of current. In the following, this time is identified as $t_s$.

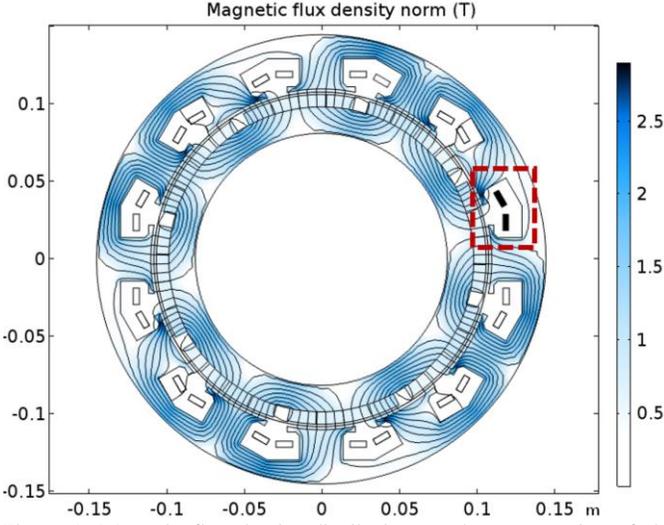

Figure 1. Magnetic flux density distribution on the cross-section of the SUTOR. The red dashed square frames one slot in which the top half-coil is called HC6 and the bottom one, HC7. The magnetic flux density is evaluated in this figure when HC7 is passed by its maximal current amplitude, at $t_s$.

*A. Application to the SUTOR motor*

For the SUTOR motor study, only the stator windings are made of REBCO, which are submitted to an alternative magnetic field while an AC current is passing through them. This causes AC losses in the windings, which need to be evaluated.

In COMSOL, the non-superconducting region is solved using the AC/DC module with Ampere law nodes, which can be computed by either $1^{st}$ or $2^{nd}$ order elements. In the meantime, the superconducting regions, i.e. the stator windings, are solved using Partial Differential Equation (PDE) nodes in which the Faraday law is implemented for the model, working with $1^{st}$ order elements.

*B. Results and validation of the model*

The model allows calculating the current distribution inside the conductor of each half-coil to assess the AC losses. The trends of the current distribution are given for HC6 and HC7 respectively in Figure 2 and Figure 3 at $t_s$, where the ratio of the current density over the critical current density, $J/J_c$ is plotted on each tape for each half-coil.

For each figure, the current distributions are evaluated with the *T-A* formulation and the MEMEP method. In this configuration, the critical current $I_c$ of the conductor is chosen to be twice the peak current passing through the conductor, i.e. 91.92 A. $I_c$ is considered constant, i.e. the field dependence of $I_c$ is neglected. The factor of the power law $n$ is 30, also considered constant.

When the *T-A* formulation is used, the conductors are modelled as a line and are enlarged in the graphs for visual purpose. In the simulation, only those two HTS half-coils are computed with the *T*-formulation and the other stator windings are simulated as copper blocks.

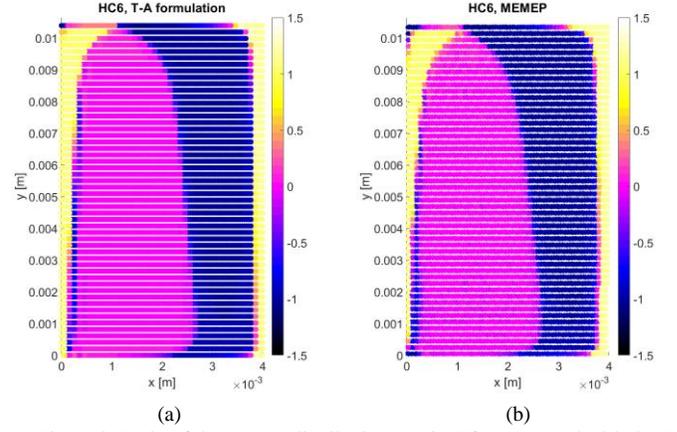

Figure 2. Ratio of the current distribution $J/J_c$ in HC6 computed with the *T-A* formulation (a) and the MEMEP method (b).

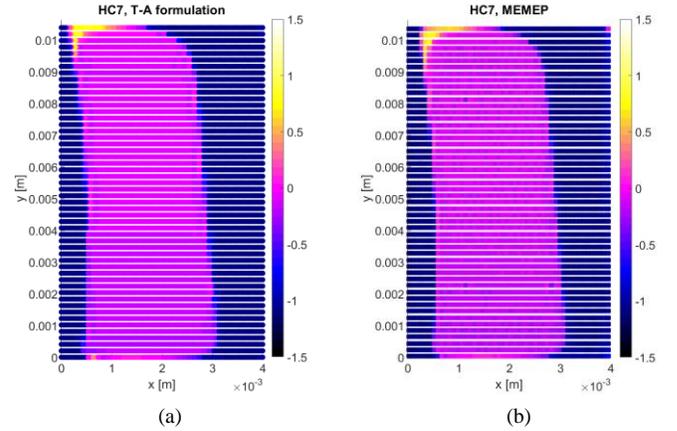

Figure 3. Ratio of the current distribution $J/J_c$ in HC7 computed with the *T-A* formulation (a) and the MEMEP method (b).

In Figure 2, the half-coil is plotted vertically even if it is not in the global view of the cross section of the motor, Figure 1, only for visual purpose. At a given time, the current density distribution in HC6 and HC7 is different as the two half-coils are not subjected to the same background field and the current going through them is also different.

The trends of the current distribution assessed with the two different methods show very good agreement for both half-coils. For a better comparison, the relative error (in %) of the values assessed using the *T-A* formulation with respect to those computed with the MEMEP method is shown in Figure 4 (a) and (b), respectively for HC6 and HC7. The relative error is defined with respect to $J_c$, and hence as

$$\text{Error (\%)} = \frac{|J_{MEMEP} - J_{TA}|}{J_c} \times 100 \qquad (5)$$

In this error definition, we use $J_c$ at the denominator instead of $J_{MEMEP}$ to avoid singularities at $J_{MEMEP} = 0$. Moreover, for all significant quantities calculated from $J$, such as local electric field, AC losses, generated $\boldsymbol{B}$ and magnetic moment, the relevant quantity is the error relative to $J_c$, or the average $|J|$ for very low current penetration.

In general, the error is only of a few %. The highest errors are in very thin sectors, at the limit between the regions with high and low current (see Figure 2 and Figure 3 for reference). In those regions, the current front is very sharp and a small difference in the position of the front in the two models creates a large error. Since these errors are limited to extremely thin regions, they have no practical consequence for the computation of quantities like the AC losses.

commercial software. The main advantage of this approach is that the simulation of the whole electrical machine and the calculation of the losses in the windings can be done within the same model. The computation times are also very reasonable: the simulation of 3 cycles typically takes one or a few hours at most on an i7-4800MQ CPU, 2.70GHz laptop computer.

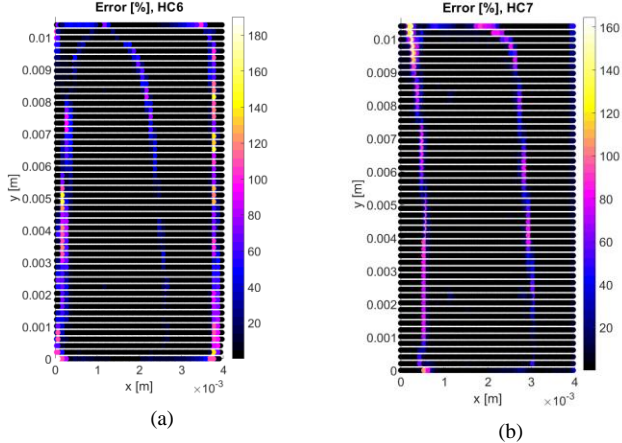

(a)    (b)

Figure 4. Relative error of the values assessed using the *T-A* formulation with respect to those computed with the MEMEP method, in HC6 (a) and HC7 (b).

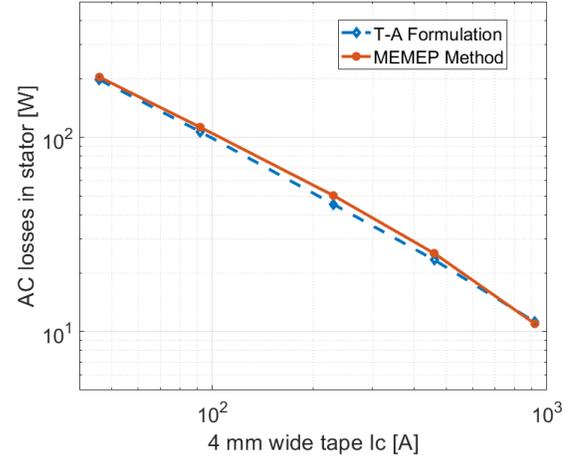

Figure 5. AC losses calculated in the stator for one cycle according to the critical current of the conductor. The results calculated from both methods agree.

The use of the *T-A* formulation to model the motor shows good results concerning the current distributions over the tapes, which allow to evaluate the AC losses and compare them to those assessed with the MEMEP. In Figure 5, the AC losses are evaluated in the stator on one cycle for different critical currents, from the value of the peak current passing through the conductor to twenty times this value. The current going through the conductors remains the same in the simulations to show the trend of the AC losses if the margins are increased, for example if the performance of the conductor is improved or if the temperature is decreased, which would imply as well a higher $I_c$.

The critical current and *n* are also considered constant in these calculations.

The AC losses computed by the two methods show similar results. The difference on the results calculated using the *T-A* formulation compared to those assessed with the MEMEP method is around 5.6 % with a maximum lower than 10 %, which confirm the previous conclusion on the current distributions. Moreover, this difference can find an explanation in the fact that with the *T-A* formulation, the complete motor is modelled whereas the MEMEP method is based on initial magnetic field distribution maps of the machine, which are calculated upstream, i.e. the direct coupling with the other parts of the motor is absent.

The study of the same motor by the *T-A* formulation and using the MEMEP method and the good results of their result comparison allow to validate that the *T-A* formulation is promising to study all kind of superconducting motors using a

## IV. ADDITION OF THE $J_c$ ANISOTROPY

In this section, $J_c$ is no longer considered constant, but dependent on the magnitude and orientation of the magnetic field. Such dependence is derived from experimental $I_c$ measurements of a SuperOx HTS coated conductor tape, which is studied in [10]. The measured conductor is a 4 mm tape, whose $I_c$ at 77 K, self-field is 168 A, and the superconducting layer is 1.8 μm thick. A symmetry over 180° is assumed to assess the $I_c$ data on the whole tape rotation.

According to the temperatures chosen for the SUTOR motor study, the simulations are carried out with the values of $J_c$ at 77 K and at 65 K [15]. These are taken from the measured values of $I_c$ in [10] simply divided by the cross-section of the superconducting layer. To match the experimental results in the model, the superconducting layer is now implemented as 1.8 μm and not 2 μm as in the previous case (for constant $J_c$).

In this case, the AC power dissipation for one cycle is respectively 52.31 W/cycle and 90.79 W/cycle at 65 K and 77 K. The AC losses computed with the MEMEP method are respectively 53.08 W/cycle and 96.18 W/cycle at 65 K and 77 K, which means that the *T-A* formulation shows a difference of 1.5 % and 5.6 %, at 65 K and 77 K. All the AC loss calculations are summarized in Figure 9. As expected, the AC losses of the stator windings are higher at 77 K than at 65 K, as the $I_c$ is lower for a given background field.

The complete 180° angular characterization of the $I_c(B)$ dependence is a lengthy process, which requires a dedicated experimental setup to rotate the sample. Measuring the $I_c(B)$ dependence at a given (fixed) field orientation is much faster and simpler. Therefore, as a second step, we repeated the AC loss calculation by considering only the dependence of $I_c$ on the field component perpendicular to the flat face of the tape. It means that in the model, only the $I_c$ value under perpendicular field are implemented and all the magnetic field applied on the conductor is assumed to be perpendicular to the tape. This is the orientation for which the reduction of $I_c$ is most significant, at least for tapes without artificial pinning centres like the ones considered here.

In this configuration, the $I_c$ values used in the model are given in Table II. In this case, the AC power dissipation for one cycle is respectively 51.76 W/cycle and 90.62 W/cycle at 65 K and 77 K, which represent a 1.1 % and 0.2 % difference from the first calculation. Using the MEMEP method, the computed AC losses are 53.63 W/cycle and 97.09 W/cycle at 65 K and 77 K, respectively, which represent a 6.5 % and 6.7 % error for the *T-A* formulation.

TABLE II
EXPERIMENTAL MEASUREMENT OF $I_C$ UNDER $B_{PERP}$

| Bperp [T] | Ic @ 65 K [A] | Ic @ 77 K [A] |
|---|---|---|
| 0 | 315 | 168 |
| 0.5 | 125 | 53 |
| 1 | 85 | 34 |
| 2 | 55 | 19 |
| 3 | 41 | 12 |

To understand the similarity on the results of the two cases, the losses are investigated for 77 K and the instantaneous dissipation is plotted in Figure 6 for the two studied half coils 6 and 7. In this slot, the main losses come from the half-coil 6 when the peak current is passing through the wound conductor.

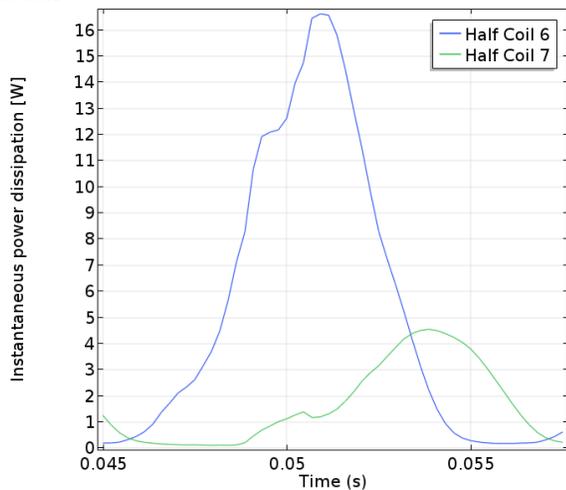

Figure 6. AC losses in the two half-coils studied at 77 K with a focus on when their peak current is passing through them.

For the peak current, the ratio of the current distribution on the $J_c$ values is plotted in Figure 7 and most of the current is passing through the left part of the coil.

The part with the highest current density on the coil creates the majority of the AC losses. Hence, in this area, the magnetic field components need to be investigated as the $J_c$ values depend on them.

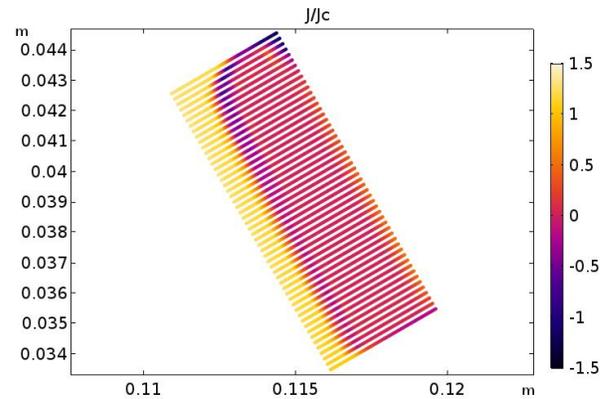

Figure 7. Normalized current density $J/J_c$ distribution inside HC6 when the peak current is passing through it at 77 K ($t$ = 0.051136 s) using all the experimental data set of $J_c$.

The magnetic flux density is plotted in Figure 8, where its intensity is given by the colour and its orientation by the arrows at the same time as in Figure 7. On the part where most of the current is passing through the conductor, i.e. the left side of the half coil, most of the field is perpendicular or almost perpendicular to the conductor. This explains why for this motor geometry study, the AC losses are similar for the case where all the $I_c$ values are implemented in the model and the case where only the $I_c$ under perpendicular field are used.

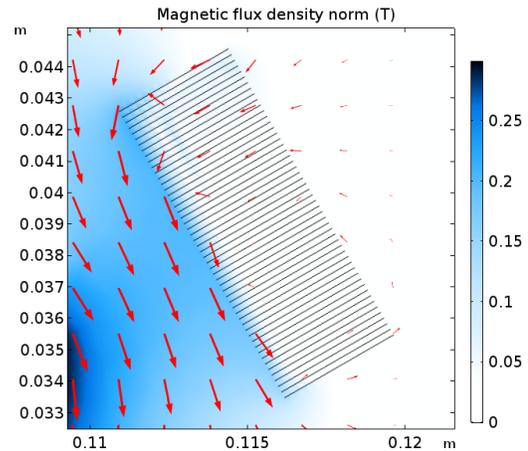

Figure 8. Magnetic field norm and orientation applied on HC6 when its peak current is passing through it at 77 K ($t$ = 0.051136 s), using all the experimental data set of $J_c$.

In the two configurations, the total AC losses are very similar, which means that for these cases and these temperatures the complete characterization of the anisotropy of $J_c$ can be neglected for this motor. In addition, the AC loss results assessed with the *T-A* formulation are similar to those computed using the MEMEP method. The difference can be explained by the fact that only the *T-A* formulation simulates the complete cross-section of the machine, as it was mentioned in the previous section.

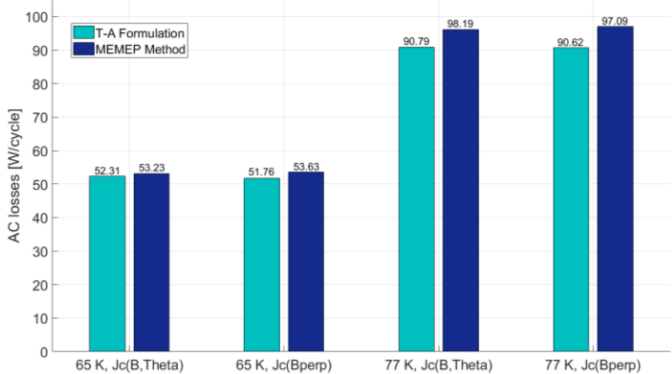

Figure 9. AC losses per cycle of the SUTOR assessed using the MEMEP method and the *T-A* formulation at 65 K and at 77 K for the cases when the anisotropy of $I_c$ is considered and when only the $I_c$ under perpendicular field are implemented.

## V. Conclusion and further work

The *T-A* formulation has already shown good results for several HTS large scale applications such as for the study of stacks of conductors, high field magnets and cables. In this paper, the *T-A* formulation is applied to simulate a motor and assess its AC losses. The results are compared with those computed with the MEMEP method, which has already been validated with analytical and experimental results, and can therefore be considered as a reliable benchmark. The *T-A* formulation shows here very good agreement for both the trend of the current distributions inside the coils as well as the AC losses over a cycle. As the *T-A* formulation is implemented using FEM, this is also the promise of the use of one commercial software to easily study AC losses of superconducting motors. After validation, the influence of the field dependence and anisotropy of the transport properties of HTS coated conductors is also studied for this motor. $J_c$ values extracted from experimental measurements are implemented in the model to show the influence of the temperature on the AC losses, which increase with the temperature due to the decrease of the $I_c$.

It has also been shown that for the analysed tape and temperature range (65 K-77 K), the whole anisotropy properties of $J_c$ can be neglected for the calculation of the AC losses, as the implementation of only the $I_c$ measured in perpendicular field is sufficient. These conclusions might not apply in other temperature ranges and with different angular field dependences.

The model is reasonably fast, but the computation speed could be improved by applying homogenization techniques for the cross-section of the superconducting coils, as it has already been done in magnets [34]. The study of different types of cable windings is also possible. Eventually, the model could also be extended to become a tool for the complete design of superconducting electrical machines.

taking account of the source current distributions and its experimental verifications," *IEEE Trans. Magn.*, vol. 27, no. 5, pp. 4020–4023, 1991.